\documentstyle[prb,aps,multicol]{revtex}

\renewcommand{\narrowtext}{\begin{multicols}{2}} 
\renewcommand{\widetext}{\end{multicols}}
\newcommand{\be}{\begin{eqnarray}}
\newcommand{\ee}{\end{eqnarray}}
\newcommand{\ba}{\begin{array}}
\newcommand{\ea}{\end{array}}
\newcommand{\no}{\nonumber}
\newcommand{\tr}{\mbox{tr}\,}
\newcommand{\str}{\mbox{str}\,}

\begin{document}

\title{Effect of spin on electron motion in a random magnetic field}
\author{K. Takahashi${}^{1}$ and K. B. Efetov${}^{1,2}$}
\address{${}^1$ Theoretische Physik III, 
 Ruhr-Universit\"at Bochum, D-44780 Bochum, Germany\\
 ${}^2$ L. D. Landau Institute for Theoretical Physics, 117940 Moscow, Russia}
\date{\today}  
\maketitle

\begin{abstract}
 We consider properties of a two-dimensional electron system 
 in a random magnetic field. 
 It is assumed that the magnetic field not only influences 
 orbital electron motion but also acts on the electron spin. 
 For calculations, we suggest the trick of 
 replacing the initial Hamiltonian by a Dirac Hamiltonian. 
 This allows us to do easily a perturbation theory and
 derive a supermatrix $\sigma$ model, 
 which takes a form of the conventional $\sigma$ model with the unitary symmetry.   
 Using this $\sigma$ model we calculate several correlation functions 
 including a spin-spin correlation function. 
 As compared to the model without spin, 
 we get different expressions for the single-particle lifetime 
 and the transport time. 
 The diffusion constant turns out to be 2 times smaller than the one
 for spinless particles. 
\end{abstract}

\narrowtext

\section{Introduction}
 The problem of electron motion in a random magnetic field (RMF) has
 attracted considerable interest in recent years. 
 One of the main questions is how the localization scenario is modified 
 in comparison with the usual disorder problem. 
 It is also thought to be relevant for the behavior of
 composite fermions near half-filling 
 in the fractional quantum Hall effect.\cite{hall} 
 In a recent experiment for a RMF system a magnetoresistance
 similar to that of the quantum Hall system was observed.\cite{exp} 

 Analytically, this problem has been discussed using 
 both the diagrammatic\cite{amw} and the supersymmetry method.\cite{amw,te1,te2} 
 It turns out that the system belongs to the usual unitary class, 
 which leads to localization in two dimensions (2D) unless 
 the random magnetic fields $B$ correlate over a very long distance 
 ($1/q^{2}$ dependence for the $\left<B_{q}B_{-q}\right>$ correlation). 
 As concerns a short range correlations of the RMF, one can derive 
 a conventional nonlinear $\sigma$ model using a standard procedure 
 (see, e.g., Ref.\onlinecite{efetov}). 
 With long range correlations one can derive first 
 a ballistic $\sigma $-model.\cite{mk} 
 Integrating out nonzero harmonics one comes
 again to the diffusive $\sigma$ model.\cite{amw,te1,te2} 
 Only if the correlation of the magnetic fields obeys the $1/q^{2}$ dependence, 
 one may get something different (antilocalization) because, in this case, 
 a new term in the $\sigma$ model appears.\cite{te2}

 A standard Hamiltonian used for the RMF problem has the form 
\be
 H_{0}=\frac{1}{2m}
 \left({\bf \hat{p}}-\frac{e}{c}{\bf A}({\bf r})\right)^{2}, \label{h0}
\ee
 where $e$ and $m$ are the electron charge and mass, and 
 the vector potential ${\bf A}({\bf r})$ corresponding to the magnetic
 field ${\bf B}\left({\bf r}\right)$ should be averaged with some weight.

 Eq.(\ref{h0}) describes electron motion in a magnetic field neglecting
 interaction of the magnetic field with the electron spin. 
 This is a good approximation for GaAs heterostructures 
 where the Zeeman splitting is very small. 
 In models of composite fermions spin is absent at all and therefore
 the Hamiltonian $H_{0}$, Eq.(\ref{h0}), is sufficient for proper description.

 Nevertheless, the question about the character of electron motion in a RMF
 acting also on the electron spin may be interesting on its own. 
 Generally, the interaction of the magnetic field with spin 
 is not smaller than interaction with the orbital motion. 
 Moreover, in a homogeneous magnetic field the Zeeman splitting for free electrons 
 is equal to the distance between Landau levels (see, e.g., Ref.\onlinecite{landau}) 
 and one may ask if this degeneracy may show up in the inhomogeneous field. 
 To the best of our knowledge the problem of electron motion 
 in a random magnetic field with both orbital and spin interaction 
 with the magnetic field has not been addressed yet.

 In this paper we address this problem starting with the Hamiltonian $H$
\be
 H=H_{0}-\frac{g}{2}\mu_B{\bf \sigma B}({\bf r}), \label{h1}
\ee
 for $g=2$.
 Here, $\mu_B=e/2mc$ is the Bohr magneton, 
 ${\bf \sigma}$ stands for the components of Pauli matrices 
 and ${\bf B}({\bf r})=\nabla\times{\bf A}({\bf r})$. 
 It is clear that one may not consider the orbital and spin interactions separately. 
 This is in contrast to models describing electron motion 
 in a magnetic field in the presence of magnetic impurities. 
 The effect of the Zeeman term has been examined in Refs.\onlinecite{mf,aa} 
 for the case of a scalar random potential and a small constant magnetic field 
 and corrections to the conductivity have been calculated.

 One should notice also that, while the interaction with the orbital motion
 is described by the vector potential ${\bf A}({\bf r})$, 
 the interaction with the spin is determined by the magnetic field 
 ${\bf B}({\bf r})$ itself. 
 Therefore, averaging over the magnetic field cannot
 be a trivial procedure and developing a proper calculational scheme may be
 interesting from the technical point of view.

 We suggest a scheme that has not been used 
 in the context of disordered metals. 
 Our idea to discuss the RMF problem for the Hamiltonian with 
 the Zeeman term is to consider a more general Dirac Hamiltonian 
 as a starting point of the analysis. 
 There are two advantages to use the Dirac Hamiltonian. 
 First, the square of the Dirac Hamiltonian gives Eq.(\ref{h1}).
 Thus, we can naturally take spin effect into consideration for the analysis.
 Second, the Dirac Hamiltonian contains only the vector potential ${\bf A}({\bf r})$ 
 but not the magnetic field ${\bf B}({\bf r})$. 
 The dependence on the gauge field ${\bf A}({\bf r})$ is linear, 
 which simplifies the averaging procedure.

 The Dirac Hamiltonian has been used for the problem of random Dirac fermions. 
 This problem may be relevant for degenerate semiconductors,\cite{fradkin} 
 quantum Hall systems,\cite{iqhe} and $d$-wave superconductors,\cite{dirac} 
 and has been under intensive study. 
 For these problems, the chiral symmetry of the Hamiltonian, 
 which means the energy eigenvalues are symmetric around the zero point, 
 plays an important role. 
 In contrast, the chiral symmetry is not important in our analysis 
 since we consider energies in the vicinity of the Fermi level and 
 therefore far from the zero energy.

 We show below that this model can be mapped onto 
 the conventional nonlinear $\sigma$ model with the unitary symmetry. 
 As the chiral symmetry is not important, this is a natural result 
 from the viewpoint of the symmetry considerations. 
 At the same time, the interaction with the spin changes the
 classical diffusion coefficient $D$ entering the $\sigma$ model. 
 We obtain somewhat different expressions for the single-particle lifetime 
 and the transport time as compared to the RMF model without spin.
 The spin degrees of freedom do not change the conventional form of 
 the current and density correlation functions. 
 As a spin dependent quantity, a spin correlation function is calculated.

\section{Self-consistent Born approximation}
 Before deriving the $\sigma$ model let us demonstrate how one can calculate
 the one-particle and transport lifetimes using the Dirac representation for
 the Hamiltonian $H$, Eq.(\ref{h1}). 
 We consider a two-dimensional system with the gauge field ${\bf A}({\bf r})$ 
 described by the Hamiltonian $H$, Eqs.(\ref{h0}) and (\ref{h1}). 
 This Hamiltonian can be represented as the square of the Dirac operator 
 $\Pi$ (we put below $c=1$) 
\be
 H &=& \frac{\Pi ^{2}}{2m} 
 = H_{0}-\mu_B\sigma_3 B_3({\bf r}), \label{h2} \\
 \Pi &=& {\bf \sigma}\left[{\bf \hat{p}}-e{\bf A}({\bf r})\right].
\ee

 We assume that the distribution of the gauge field ${\bf A}({\bf r})$ is
 Gaussian with the correlations 
\be
 \left< A_{i}({\bf r})A_{j}({\bf r}^{\prime})\right> &=& 
 \frac{2m^2}{e^2}\int\frac{d^2{\bf q}}{(2\pi)^2}V_{ij}({\bf q})
 e^{i{\bf q}({\bf r}-{\bf r}^{\prime})},   \\
 V_{ij}({\bf q}) &=&\frac{v_F^2\gamma}{q^2+\kappa^2p_F^2}
 \left(\delta_{ij}-\frac{q_iq_j}{q^2}\right),  
\ee
 where $\gamma$ characterizes the strength of the disorder 
 and $p_F$ is the Fermi momentum. 
 The limit $\kappa=0$ corresponds to a $\delta$-correlated magnetic field. 
 Nonzero values of $\kappa$ describe a screening
 and we keep below an arbitrary value of $\kappa$.

 Our main idea for considering the spin effects in the random magnetic field
 problem is to represent Green functions $G_E^{R,A}$ 
\be
 G_E^{R,A}=\frac{1}{E-H\pm i\delta} 
\ee
 of the Hamiltonian $H$, Eqs. (\ref{h0}) and (\ref{h1}), 
 in terms of the Green functions $g_{k}^{R,A}$ 
 of the Dirac Hamiltonian $\Pi$ 
\be
 g_{k}^{R,A} = \frac{1}{k-\Pi\pm i\delta}. 
\ee
 Using the relation 
\be
 G_{E}^{R,A}=\frac{m}{k}\left(g_{k}^{R,A}-g_{-k}^{A,R}\right), \label{gf1}
\ee
 where $k=(2mE)^{1/2}$ is actually the Fermi momentum, we achieve this goal.

 The Dirac Hamiltonian $\Pi$ is linear in the gauge field 
 ${\bf A}({\bf r})$ and the ensemble averaging can be performed easily. 
 On the other hand, the usual Hamiltonian (\ref{h0}) includes the square of the
 gauge field, which makes the averaging
 procedure more difficult. 
 At the same time, just neglecting the term ${\bf A}^{2}$ 
 may be dangerous because this violates the gauge invariance. 
 In our case, we can keep the gauge invariance at any step of calculations.

 In order to demonstrate how our scheme works we calculate first the average
 Green functions $\left<g_k^{R,A}\right>$ 
 using the well known self-consistent Born approximation. 
 Summing only ladder diagrams in the standard way\cite{agd} 
 we obtain for the Green function 
\be
 \left<g_{k}^{R,A}({\bf p})\right> =\frac{1}{k\pm i\delta 
 -{\bf \sigma p}+\Sigma^{R,A}({\bf p})},  \label{gf2}
\ee
 where the self-energy $\Sigma \left( {\bf p}\right) $ should be found from
 the equation 
\be
 \Sigma^{R,A}({\bf p}) &=& 2im^2\sum_{i,j}
 \int\frac{d^2{\bf q}}{(2\pi)^2}V_{ij}({\bf p}-{\bf q})  \no\\
 & & \times \sigma_i\frac{1}{k\pm i\delta-{\bf \sigma q}
 +\Sigma^{R,A}({\bf q})}\sigma_j.  \label{se} 
\ee
 In Eqs.(\ref{gf2}) and (\ref{se}), ${\bf p}$ and ${\bf q}$ are momenta. 
 Using the usual assumption that disorder is weak one can obtain easily 
\be
 \left<g_k^{R,A}({\bf p})\right> &=& \frac{1}{k-{\bf \sigma p}
 \pm\frac{i}{8E_F\tau}(k+{\bf \sigma p})} \label{sp} \\
 &\sim& \frac{1}{2m}\frac{k+{\bf \sigma p}}{E-\frac{{\bf p}^2}{2m}
 \pm \frac{i}{2\tau}},  
\ee
 where $\tau$ is the single-particle lifetime specified below. 
 Using Eq.(\ref{gf1}), we extract the Green function 
\be
 \left<G_E^{R,A}({\bf p})\right> = 
 \frac{1}{E-\frac{{\bf p}^2}{2m}\pm\frac{i}{2\tau}}.  \label{G}
\ee
 This is the usual form of the Green functions. 
 So, we conclude that the interaction of the magnetic spin 
 does not change the form of the Green functions.

 At the same time, the expression of the single-particle lifetime $\tau$
 differs from that for spinless particles. 
 For short-range disorder $\kappa \gg (E_{F}\tau )^{-1}$  
 the solution of Eq.(\ref{se}) leads to the following expression 
 for the single particle lifetime $\tau$: 
\be
 \frac{1}{\tau} &=&\gamma E_F\int_0^{2\pi}\frac{d\theta}{2\pi}
 \frac{1}{\sin^2\frac{\theta}{2}+\frac{\kappa^2}{4}}  \label{ba} \\
 &\sim& \frac{2\gamma E_F}{\kappa}.
\ee
 In contrast to the corresponding result 
 for the model without spin\cite{amw,te2} 
 a factor $\cos^2(\theta/2)$ in the integrand is absent.
 However, in the limit of small $\kappa$, integration over $\theta$ 
 in Eq.(\ref{ba}) leads to the same result as for the spinless problem.

 In the limit of long-range disorder $\kappa \ll (E_{F}\tau )^{-1}$, 
 we obtain solving Eq.(\ref{se}) a more complicated expression 
\be
 \frac{1}{\tau} &=& \frac{2\gamma E_F\tau}{\pi}\int_{-\infty }^{\infty}
 d\xi\int_{0}^{2\pi}\frac{d\theta}{2\pi}\sin^2\frac{\theta}{2} \no \\
 & & \times \frac{1}{\left(\frac{\xi^2}{16E_F^2}+\sin^2\frac{\theta}{2}
 +\frac{\kappa^2}{4}\right) \left(\frac{\xi^2}{16E_F^2}
 +\sin^2\frac{\theta}{2}\right)}  \label{scba} \\
 &\sim& 4E_F\left(\frac{\gamma}{\pi}\ln\frac{2}{\kappa}\right)^{1/2}.
\ee
 Again, the integrand in Eq.(\ref{scba}) differs from the corresponding
 integrand of the spinless problem by the absence of $\cos^2(\theta/2)$ 
 in the integrand. 
 This changes the final result and 
 we obtain a single-particle lifetime $\tau$ which has an additional factor
 $1/\sqrt{2}$ as compared to the corresponding result for the spinless problem.

 Due to the long range correlations of the random field 
 the transport time $\tau_{\rm tr}$ entering 
 the classical diffusion coefficient $D$ differs from
 the single particle time $\tau$. 
 By considering the nonsingular corrections as explained in Ref.\onlinecite{amw} 
 one comes to the renormalization 
\be
 \frac{1}{\tau_{\rm tr}}=\frac{1}{\tau}-\frac{1}{\tau^{(1)}}, \label{tautr1}
\ee
 where $\tau^{(1)}$ can be obtained by inserting the additional factor 
 $\cos\theta$ in the integrands in Eqs.(\ref{ba}) and (\ref{scba}). 
 For the model without spin one has $1/\tau_{\rm tr}=\gamma E_F$\cite{amw} and 
 the diffusion constant is given by $D=v_F^2\tau_{\rm tr}/2$. 
 In the present case, one comes to the following expression for 
 $1/\tau_{\rm tr}$ [again, the factor $\cos^{2}\left(\theta/2\right)$ 
 is absent in the integrand] 
\be
 \frac{1}{\tau_{\rm tr}} &=&\gamma E_{F}\int_{0}^{2\pi }\frac{d\theta }{2\pi }%
 \frac{1}{\sin ^{2}\frac{\theta }{2}}(1-\cos \theta )  \nonumber \\
 &=& 2\gamma E_{F}.  \label{tautr2}
\ee
 The diffusion constant $D$ is related to transport time 
 $\tau_{\rm tr}$ in the usual way 
\be
 D=\frac{v_F^2\tau_{\rm tr}}{2}. \label{dc}
\ee
 This result shows that by taking into account spin, 
 we get a transport time which is one half of the usual one. 
 This difference can be considered as a spin effect. 
 Remarkably, the orbital and spin scattering give equal
 contributions to the resistivity determined by $1/\tau_{\rm tr}$. 
 This effect is not trivial and is obtained only after 
 integration over $\theta $ in Eq.(\ref{tautr2}) and 
 in the corresponding equation for the spinless problem. 
 It is specific for two dimensions.

\section{Supersymmetric nonlinear $\sigma$ model}
 In order to consider interference effects and localization 
 one should consider either more complicated diagrams 
 or derive a nonlinear $\sigma$ model. 
 Both the methods have been used for the spinless problem.\cite{wb,amw,te2} 
 For the present problem we want to use the supersymmetry technique 
 and to derive a supermatrix $\sigma$ model. 
 We use the notation and conventions of Ref.\onlinecite{efetov} 
 in the following calculation.

 As we have seen, we need the Green functions $g_{k}^{R,A}$ and $g_{-k}^{R,A}$
 in order to calculate the Green functions $G_{E}^{R,A}$. 
 We define the following generating function to calculate 
 the product of the Green's function as 
\be
 Z(k,\tilde{\omega}) &=&\int D\psi D\bar{\psi}\,\mbox{e}^{-{\cal L}}, \\
 {\cal L} &=&-i\int d^{2}{\bf r}\,\bar{\psi}({\bf r})\left[H_{0}+e\alpha 
 {\bf \Sigma}{\bf A}({\bf r})\right] \psi({\bf r}),
\ee
 where 
\be
 H_0 = -i\alpha{\bf \sigma\nabla}-k+\frac{\tilde{\omega}^+}{2}\Lambda. 
\ee
 $k$ denotes the Fermi {\it momentum} and $\tilde{\omega}$ is related to 
 the energy difference as $\omega =v_F\tilde{\omega}$. 
 The matrices $\Sigma$ and $\alpha$ are defined as 
\be
 \Sigma_{i} &=& \sigma_{i}\tau_{3} = 
 \pmatrix{ \sigma_i & 0 \cr 0 & -\sigma_i \cr}, \\
 \alpha &=&\pmatrix{ 1 & 0 \cr 0 & -1 \cr}.
\ee
 Note that ${\bf \sigma}$ are the (Pauli) matrices in the spin space, 
 $\tau_{3}$ in the time-reversal space and $\alpha$ is a matrix that reflects 
 $+\Pi $ and $-\Pi$. 
 The supervector $\psi$ contains 32 components corresponding to 
 fermion/boson, advanced/retarded, time-reversal multiplication, 
 spin degrees of freedom, and $+$/$-$ structure.

 The calculation is done in a similar way as that for the model without spin.
 After the ensemble averaging and the Hubbard-Stratonovitch transformation,
 the generating function can be written as 
\widetext
\be
 \left<Z(k,\tilde{\omega})\right> &=& \int DQ\exp \left[
 \frac{1}{2}\,\mbox{str}\,\ln \left(iH_{0}\delta({\bf r}-{\bf r}^{\prime})
 -2m^2\sum_{i,j}V_{ij}({\bf r}-{\bf r}^{\prime })\alpha \Sigma _{i}Q({\bf r},{\bf r}%
 ^{\prime })\alpha \Sigma _{j}\right)\right.  \no \\
 & & \left. -\frac{m^2}{2}\sum_{i,j}\int d^2{\bf r}d^2{\bf r}^\prime
 V_{ij}({\bf r}-{\bf r}^\prime)
 \str\,Q({\bf r},{\bf r}^\prime)\alpha\Sigma_i
 Q({\bf r}^\prime,{\bf r})\alpha\Sigma_j\right],
\ee
\narrowtext 
 where $Q$ is a 32$\times$32 supermatrix and ``str'' denotes the supertrace.

 The saddle-point equation is solved by 
\be
 Q^{(0)}({\bf p})=\frac{i}{\alpha {\bf \Sigma p}-k+\frac{i}{8E_{F}\tau}
 (k+\alpha {\bf \Sigma p})\Lambda},  \label{sp2}
\ee
 which is essentially the same as Eq.(\ref{sp}). 
 The single-particle lifetime $\tau$ has already been obtained above.

 Now let us turn to the discussion of the saddle-point manifold. 
 For $\omega=0$, the generating function is invariant under 
 the transformation $\psi\rightarrow T\psi$, 
 where the matrix $T$ obeys the following constraints:
\be
 T\bar{T} &=& 1, \\
 T\alpha\Sigma_{1,2}\bar{T} &=& \alpha\Sigma_{1,2}.
\ee
 The first equation is the unitarity condition, whereas the second one
 specifies the structure in the spin space. 
 Using these conditions we obtain the saddle-point manifold. 
 It can be parametrized as 
\be
 Q &=& TQ^{(0)}\bar{T}, \label{spm1}\\ 
 T &=& \left(\frac{1-iP}{1+iP}\right)^{1/2}, \label{spm2}\\
 P &=& \pmatrix{ 0 & B \cr \bar{B} & 0 \cr}, \quad
 B = \pmatrix{ B^{(++)} & B^{(+-)} \cr B^{(-+)} & B^{(--)} \cr}. 
\ee
 This block form of the matrix $B$ reflects the structure in $+$/$-$ space.
 This matrix satisfies $[B,\alpha\Sigma_{1,2}]=0$ in addition to the usual
 condition $B=K\bar{B}K$. 
 It means that $B^{(++)}$ and $B^{(--)}$ are proportional to 
 the unit matrix in spin space 
 and $B^{(+-)}$ and $B^{(-+)}$ are proportional to $\Sigma _{3}$. 
 This result shows that if we consider the generating function 
 for the Dirac Hamiltonian $\Pi$, 
 only the mode $B^{(++)}$ is left gapless 
 and we come to the usual unitary $\sigma$ model. 
 The structure in $+$/$-$ space makes the result more complicated.
 We have four gapless modes and they equally contribute to 
 the physical quantities as we see below.
 
 Having specified the saddle-point manifold, 
 we derive the free energy describing gapless modes. 
 Again, this calculation resembles that for the spinless case. 
 In addition to the usual massless mode $P({\bf R})$, 
 we need to take into account soft massive modes $P({\bf R},{\bf n})$ 
 which depend on the direction vector of the Fermi momentum 
 (${\bf n}={\bf p}/|{\bf p}|)$. 
 The saddle-point manifold is parametrized as Eq.(\ref{spm1})
 and the matrix $T$ is decomposed as 
\be
 T({\bf R},{\bf n}) = U({\bf R})V({\bf R},{\bf n}). 
\ee
 The supermatrices $U$ and $V$ parametrize the massless modes and 
 the soft massive modes respectively. 
 They are expressed as Eq.(\ref{spm2})
 and expanded in the supermatrix $P$.
 For the soft massive modes, the dependence of the supermatrices on ${\bf n}$ 
 is treated in the Fourier-transformed space as 
\be
 P({\bf R},{\bf n})=\sum_{m}P({\bf R},m)\mbox{e}^{im\theta\tau_3}, 
\ee
 where $\theta$ is the polar angle of the vector ${\bf n}$. 
 The $m$th lifetime $\tau^{(m)}$ which is associated with each modes is defined as 
\be
 & & \frac{k}{8E_{F}\tau ^{(m)}}(1+\alpha {\bf \Sigma n)}\Lambda \no\\
 &=& 2m^{2}\sum_{i,j}\int\frac{d^{2}{\bf q}}{(2\pi)^2}
 V_{ij}({\bf p}-{\bf q})\alpha\Sigma_{i}Q^{(0)}({\bf p})
 \alpha\Sigma_{j}\mbox{e}^{im\theta}.  \no\\ 
 & &
\ee
 $\tau^{(0)}$ corresponds to the single-particle lifetime.
 We will see that these nonzero-harmonics modes renormalize the transport time 
 and, as a result, one comes to Eq.(\ref{tautr1}). 
 Details are presented in Ref.\onlinecite{te2}.
 Performing the gradient expansion and integrating out the soft massive modes
 we obtain the nonlinear $\sigma$ model 
\be
 F &=& \frac{\pi\nu}{16}\int d^2{\bf r}\,\str 
 \left[ D[{\bf \nabla}Q({\bf r})]^2+2i\omega\Lambda Q({\bf r})\right], 
\ee
 where $Q({\bf r})=U({\bf r})\Lambda\bar{U}({\bf r})$ and 
 $\nu =m/2\pi$ is the density of states. 
 The diffusion constant $D$ is given by Eq.(\ref{dc}) and 
 the transport time $\tau_{\rm tr}$ is introduced in Eq.(\ref{tautr2}).

 The form of the free energy $F$ is the same as for the model without spin
 but the dependence of the diffusion coefficient $D$ on the correlations of
 the magnetic field is somewhat different. 
 We note that an additional term derived in Ref.\onlinecite{te2} 
 can also be present in our model. 
 However, this term appears at longer correlations of the magnetic field 
 and we will not consider it in this paper.

 Due to the spin degrees of freedom, the $32\times 32$ supermatrix $Q$ 
 has a more complicated structure than usual. 
 However, we will show below, the renormalization properties show 
 this model falls into the usual unitary class. 
 We show the results for the conductivity and the current correlation function. 
 They can be calculated by using the following contraction rules
 for the perturbative calculations as 
\widetext
\be
 \left<\str XP({\bf r})YP({\bf r}^{\prime})\right>_F
 &=& \frac{1}{16}\Pi({\bf r},{\bf r}^\prime)\sum_{\mu=0}^{3}
 [\str\sigma_{\mu}X\str\sigma_{\mu}Y
 -\str\Lambda\sigma_{\mu}X\str\Lambda\sigma_{\mu}Y \no \\
 & & +\str\alpha\sigma_{\mu}X\str\alpha\sigma_{\mu}Y
 -\str\alpha\Lambda\sigma_{\mu}X\str\alpha\Lambda\sigma_{\mu }Y \no\\
 & & +\str\sigma_{\mu}\sigma_{3}X\str\sigma_{\mu}\sigma_{3}Y
 -\str\Lambda\sigma_{\mu}\sigma_{3}X\str\Lambda\sigma_{\mu }\sigma_{3}Y \no\\
 & & -\str\alpha\sigma_{\mu}\sigma_{3}X\str\alpha\sigma_{\mu}\sigma_{3}Y
 +\str\alpha\Lambda\sigma_{\mu}\sigma_{3}X\str\alpha\Lambda\sigma_{\mu}\sigma_{3}Y \no \\
 & & +(X\rightarrow\tau_{3}X, Y\rightarrow\tau_{3}Y) ], \\
 \left<\str XP({\bf r})\str YP({\bf r}^{\prime})\right>_F 
 &=& \frac{1}{8}\Pi({\bf r},{\bf r}^\prime)\sum_{\mu=0}^3
 \str (\sigma_{\mu}X\sigma_{\mu}Y-\Lambda\sigma_{\mu}X\Lambda\sigma_{\mu}Y
 +\alpha\sigma_{\mu}X\alpha\sigma_{\mu}Y
 -\alpha\Lambda\sigma_{\mu}X\alpha\Lambda\sigma_{\mu}Y  \no \\
 & & +\sigma_{\mu}\sigma_3X\sigma_\mu\sigma_3Y
 -\Lambda\sigma_{\mu}\sigma_{3}X\Lambda\sigma_{\mu}\sigma_{3}Y
 -\alpha\sigma_{\mu}\sigma_{3}X\alpha\sigma_{\mu}\sigma_{3}Y
 +\alpha\Lambda\sigma_{\mu}\sigma_{3}X\alpha\Lambda\sigma_{\mu}\sigma_{3}Y) \no\\
 & & +(Y\rightarrow -\bar{Y}),
\ee
\narrowtext
 where $X$ and $Y$ are arbitrary 32$\times $32 matrices which commute with 
 $\tau_{3}$ and $\Pi(x,y)$ is the diffusion propagator 
\be
 \Pi({\bf r},{\bf r}^\prime) = \frac{1}{2\pi\nu}
 \int\frac{d^2{\bf q}}{(2\pi)^{2}}
 \frac{1}{D{\bf q}^2-i\omega}\mbox{e}^{i{\bf q}({\bf r}-{\bf r}^{\prime})}. 
\ee
 At first glance, the above expressions of the contraction rules look quite
 different from the usual ones. 
 However, if we assume that $X$ and $Y$ do not have any spin structure, 
 we obtain the standard contraction rules.
 Therefore, we will get the usual unitary results 
 for spin-independent quantities. 
 We show for instance the results of the conductivity and 
 the current correlation function. 
 The conductivity is defined as 
\be
 \sigma_{ij}(\omega)=\frac{e^{2}}{2\pi}R_{ij}({\bf k}=0,\omega), 
\ee
 where $R_{ij}({\bf k},\omega)$ is the Fourier-transform of the correlation function 
\be
 R_{ij}({\bf r},{\bf r}^{\prime },\omega) &=& \left< \tr
 \hat{\pi}^i_{\bf r}G_{E-\omega/2}^{A}({\bf r},{\bf r}^{\prime })
 \hat{\pi}^j_{{\bf r}^{\prime}}G_{E+\omega /2}^{R}({\bf r}^{\prime},{\bf r})
 \right>,\no\\
 & &
\ee
 where ${\bf \hat{\pi}}_{{\bf r}}=-i{\bf \nabla}_{{\bf r}}/m$ is 
 the velocity operator. 
 In the two-loop approximation we obtain 
\be
 \sigma_{ij}(\omega) &=& \sigma_0\delta_{ij}\left[
 1+\frac{1}{2\pi^2\nu^2}\left(1-\frac{2}{d}\right) 
 \left(\int\frac{d^{d}{\bf q}}{D{\bf q}^{2}-i\omega}\right)^{2}\right], \no\\
 & & 
\ee
 where $\sigma_0=2e^2\nu D$ is the classical conductivity. 
 This is the same expression as the one presented in Ref.\onlinecite{te2} and 
 is just the usual unitary result.

 The current correlation function which is defined as 
\be
 I_{ij}({\bf r},{\bf r}^{\prime},\omega) &=& \left<J_{i}({\bf r},E-\omega/2)
 J_{j}({\bf r}^{\prime},E+\omega/2)\right>, \\
 J_{i}({\bf r},E) &=& \frac{ie}{4\pi}\lim_{{\bf r}^{\prime}\to{\bf r}}
 \left(\hat{\pi}^i_{{\bf r}}-\hat{\pi}^i_{{\bf r}^{\prime}}\right) \no \\
 & & \times \tr\left[G_{E}^{R}({\bf r},{\bf r}^{\prime})
 -G_{E}^{A}({\bf r},{\bf r}^{\prime})\right],
\ee
 can be written in a Fourier-transformed form in the leading order as 
\be
 I_{ij}({\bf q},\omega) = \frac{2e^2}{\pi^3}
 \left(\delta_{ij}-\frac{q_{i}q_{j}}{q^{2}}\right) 
 \ln\left(qL_{\omega}\right),  
\ee
 where $L_{\omega}=(D/\omega)^{1/2}$. 
 This result is four times larger than the corresponding result 
 of Ref.\onlinecite{gmw}. 
 This difference is due to the spin degrees of freedom. 
 We see that all the expressions for the spinless quantities are just 
 the results for the unitary ensemble.

 For the present model, we can introduce in addition spin correlation
 functions that could not exist for models without spin. 
 As an example, we define a correlation function which is a direct extention 
 of the density-density correlation function widely used 
 in theory of localization\cite{efetov} 
\be
 T_{ij}({\bf q},\omega) &=& -i\int\frac{dE}{2\pi}
 \int\frac{d^2{\bf p}}{(2\pi)^{2}}\biggl[\left(n(E)-n(E-\omega)\right) \no\\
 & & \times\left< \tr \sigma_{i}G^{R}_E({\bf p})
 \sigma_{j}G^{A}_{E-\omega}({\bf p}-{\bf q})\right>  \no \\
 & & +n(E)\left< \tr\sigma_{i}G^{R}_{E+\omega}({\bf p}+{\bf q})
 \sigma_{j}G^{R}_E({\bf p})\right. \no\\
 & & \left. -\tr\sigma_{i}G^{A}_E({\bf p})
 \sigma_{j}G^{A}_{E-\omega}({\bf p}-{\bf q})\right> \biggr], \label{D}
\ee
 where $n(E)$ is the Fermi distribution function and 
 $i,j=0,1,2,3$ ($\sigma_0=1$). 
 The component $T_{00}({\bf q},\omega)$ gives 
 the conventional density-density correlation function. 
 We refer to Appendix for details of the calculation and obtain 
\be
 T_{ij}({\bf q},\omega) &=& \left\{\ba{cc}
 2\nu\frac{D{\bf q}^2}{D{\bf q}^2-i\omega} & \mbox{for $i=j=0,3$}, \\ 
 2\nu & \mbox{for $i=j=1,2$}.
 \ea\right. \label{Dr}
\ee
 We find the usual result for the density correlation function.
 Since we do not take any electron interaction into account, 
 this can be considered a natural result.
 The same diffusive form appears for $T_{33}$.
 This diffuson contribution comes from the mixing of the $+$/$-$ space.

\section{Conclusion}
 In conclusion, we have discussed spin effects 
 for the two-dimensional random magnetic field model. 
 We suggested to take into account spin degrees of freedom 
 writing the Dirac Hamiltonian instead of the initial Hamiltonian 
 for electrons with spin, which is possible for the model of free electrons. 
 The interaction of the magnetic field with spin is not smaller than its effect
 on the orbital motion. 
 Moreover, in a homogeneous field the Zeeman splitting
 is equal to the distance between the Landau levels.

 We derived a nonlinear supermatrix $\sigma$ model that turned out to be 
 a conventional $\sigma$ model with the unitary symmetry. 
 However, the single particle lifetime and the diffusion coefficient 
 differ from their values for the spinless particles. 
 Remarkably, the transport time and, correspondingly, 
 the diffusion coefficient are two times smaller than those 
 for the spinless problem. 
 This means that the orbital and spin scattering equally contribute to resistivity. 
 The conductivity and the current-current correlation functions 
 take the conventional form for the unitary class. 
 The form of the spin-spin correlation function is the similar to 
 that of the density-density correlation function. 
 Our results obtained by the supersymmetry method can be well reproduced 
 using the diagrammatic methods.

 Finally, we mention two possible extentions of our model.
 First, our analysis is restricted to the model with fixed g factor $g=2$.
 The advantage to introduce the Dirac Hamiltonian 
 was that the spin degrees of freedom is taken into account naturally.
 The square of the Dirac Hamiltonian without any parameter 
 gives Eq.(\ref{h2}) with fixed g factor.
 On the other hand, it is well known that 
 the g factor can take different value for realistic materials.
 Unfortunately, it is difficult to extend our approach  
 directly to the case of a different g factor.
 However, we hope that our results shed some light on a possible behavior
 of more realistic systems.

 Second interesting problem is the case of 
 a nonzero average magnetic field.
 In this case, the Zeeman splitting causes 
 the reformation of the Landau levels and nontrivial results may be expected. 
 For spinful electrons, in addition to the perpendicular magnetic field, 
 the parallel field is also relevant even for the two-dimensional system.
 In this paper, we have discussed a two-dimensional Hamiltonian. 
 The Dirac Hamiltonian does not include the Pauli matrix $\sigma_{3}$ 
 and only the perpendicular magnetic field enters 
 the square of the Dirac Hamiltonian [Eq.(\ref{h2})]. 
 If we consider the three-dimensional Hamiltonian, 
 we can include the magnetic field for all directions. 
 Even if the diffusion process is two-dimensional, 
 we may need to treat the one-electron states 
 as three-dimensional in that case. 
 It will be a subject of a future work.

\section*{Acknowledgments}
 We acknowledge the financial support of the Sonderforschungsbereich 237.
 K.T would like to thank G. Schwiete for discussions and 
 reading the manuscript.

\appendix
\section{Calculation of the spin correlation function} 
 We consider the function 
\be
 \left< \tr \sigma_{i}G^{R}_E({\bf r},{\bf r}')
 \sigma_{j}G^{A}_{E-\omega}({\bf r}',{\bf r})\right>,
\ee
 to calculate the spin correlation function.
 For the product of the Green functions,
 we express it in the functional integral form as 
\be
 & & \left< G^{(A)\beta\alpha}_{E-\omega}({\bf r},{\bf r}')
 G^{(R)\delta\gamma}_{E}({\bf r}',{\bf r}) \right> \no\\
 &=& -\frac{4}{v_F^2}\int D(\psi\bar{\psi})
 [\psi^{1+}_{\mu\beta}({\bf r})\bar{\psi}^{1+}_{\mu\alpha}({\bf r})
 +\psi^{1-}_{\mu\beta}({\bf r})\bar{\psi}^{1-}_{\mu\alpha}({\bf r})] \no\\
 & & \times 
 [\psi^{2+}_{\nu\delta}({\bf r}')\bar{\psi}^{2+}_{\nu\gamma}({\bf r}')
 +\psi^{2-}_{\nu\delta}({\bf r}')\bar{\psi}^{2-}_{\nu\gamma}({\bf r}')]
 \mbox{e}^{-{\cal L}},
\ee
 where  $\alpha$, $\beta$, $\gamma$, and $\delta$ are spin indices, 
 1 and 2 denote the advanced and retarded channel, 
 $+$ and $-$ the $+$/$-$ space, 
 and $\mu$ and $\nu$ are other indices.
 No summation is implied for $\mu$ and $\nu$.
 After introducing the auxiliary field $Q$ and 
 taking the contraction of $\psi$ in two possible ways, we get 
\be
 & & \frac{1}{v_F^2}
 \biggl<(\sigma_i)_{\alpha\beta}(\sigma_j)_{\gamma\delta}
 \str \Bigl[ C_+ g_{\beta\alpha}({\bf r},{\bf r})
 C_- g_{\delta\gamma}({\bf r}',{\bf r}')\Bigr]\biggr>  \no\\
 & & -\frac{1}{v_F^2}
 \biggl< (\sigma_i)_{\alpha\beta}(\sigma_j)_{\gamma\delta}
 \str \Bigl[ C_+ g_{\beta\gamma}({\bf r},{\bf r}')\Bigr] 
 \str \Bigl[ C_- g_{\delta\alpha}({\bf r}',{\bf r})\Bigr]\biggr>,  \no\\ \label{I}
\ee
 where 
\be
 C_{\pm} &=& \frac{k}{2}\frac{1-\tau_3}{2}\frac{1\pm\Lambda}{2}.
\ee
 At the leading order, the Green's function $g$ is given 
 in the Fourier-transformed space in terms of the relative coordinates as
\be
 g({\bf R},{\bf p}) &=& \frac{i}{\alpha{\bf \Sigma} {\bf p}-\epsilon
 +\frac{i}{8E_F\tau}(\epsilon+\alpha{\bf \Sigma} {\bf p})Q({\bf R})}, 
\ee
 and $Q({\bf R})$ is expanded in terms of $P({\bf R})$.

 Substituting $g$ to the first term of Eq.(\ref{I}), we have leading contribution
\be
 \pi^2\nu^2
 \biggl<(\sigma_i)_{\alpha\beta}(\sigma_j)_{\gamma\delta}
 \str \Bigl[ C_+ P_{\beta\alpha}({\bf r})
 C_- P_{\delta\gamma}({\bf r}')\Bigr]\biggr>. 
\ee
 The matrix $P$ has the structure 
\be
 P &=& \left(\ba{cc} P^{(++)} & P^{(+-)} \\ P^{(-+)} & P^{(--)} \ea\right),
\ee
 in $+$/$-$ space.
 $P^{(++)}$ and $P^{(--)}$ are proportional to $\sigma_0$ in spin space and
 $P^{(+-)}$ and $P^{(-+)}$ are proportional to $\sigma_3$.
 Using this fact and the contraction rule, we get
\be
 & & 2\pi^2\nu^2\biggl[
 \delta_{i0}\delta_{j0}\Bigl<
 \str C_+P^{(++)}({\bf r})C_-P^{(++)}({\bf r}') \no\\
 & & +\str C_+P^{(--)}({\bf r})C_-P^{(--)}({\bf r}')\Bigr> \no\\
 & & +\delta_{i3}\delta_{j3}
 \Bigl<\str C_+P^{(+-)}({\bf r})C_-P^{(-+)}({\bf r}') \no\\
 & & +\str C_+P^{(-+)}({\bf r})C_-P^{(+-)}({\bf r}')\Bigr>
 \biggr] \no\\
 &=&  4\pi\nu
 \left(\delta_{i0}\delta_{j0}+\delta_{i3}\delta_{j3}\right)
 \int\frac{d^2{\bf q}}{(2\pi)^2}
 \frac{1}{D{\bf q}^2-i\omega}\mbox{e}^{i{\bf q}({\bf r}-{\bf r}')}.
\ee
 The contribution comes from the second term of Eq.(\ref{I}) is small and is neglected.

 We use this result for the first term of Eq.(\ref{D}). 
 Substituting the Green's function (\ref{G}) to 
 the second term of Eq.(\ref{D}), we find $2\nu\delta_{ij}$. 
 Combining these results, we finally obtain Eq.(\ref{Dr}). 


\widetext

\end{document}